\documentstyle[prl,aps,multicol,amssymb,amsfonts]{revtex}

\def \be#1\ee {\begin{equation}#1\end{equation}}
\newcommand{\ds}{\displaystyle}
\newcommand{\br}{{\bf r}}

\renewcommand{\Re}{\mathop{\rm Re}\nolimits}
\renewcommand{\Im}{\mathop{\rm Im}\nolimits}

\newcommand{\str}{\mathop{\rm str}\nolimits}
\newcommand{\const}{{\rm const}}
\newcommand{\gap}{E_{\rm g}}
\newcommand{\ETh}{E_{\rm Th}}
\newcommand{\K}{\tilde{G}}
\newcommand{\dos}{\left<\rho(E)\right>}

\newcommand{\sx}{\sigma_x}
\newcommand{\sy}{\sigma_y}
\newcommand{\sz}{\sigma_z}
\newcommand{\tx}{\tau_x}
\newcommand{\ty}{\tau_y}
\newcommand{\tz}{\tau_z}

\newcommand{\vphi}{\varphi}
\newcommand{\eps}{\varepsilon}

\newcommand{\fb}{{\rm FB}}

\begin{document}

\title{Density of States below the Thouless Gap
in a Mesoscopic SNS Junction}

\author{P. M. Ostrovsky, M. A. Skvortsov, and M. V. Feigel'man}

\address{L. D. Landau Institute for Theoretical Physics,
117940 Moscow, Russia}


\maketitle

\begin{abstract}
Quasiclassical theory predicts an existence of a sharp energy gap $\gap
\sim \hbar D/L^2$ in the excitation spectrum of a long diffusive
superconductor--normal metal--superconductor (SNS) junction.
We show that mesoscopic fluctuations remove the sharp edge of the
spectrum,
leading to a nonzero DoS for all energies.
Physically, this effect originates from the quasi-localized states
in the normal metal. Technically, we use an extension of Efetov's
supermatrix $\sigma$-model for mixed NS systems.
A non-zero DoS at energies $E < \gap$
is provided by the instanton solution with broken supersymmetry.
\end{abstract}

\begin{multicols}{2}

When a small piece of a normal (N) metal is placed in contact with a
superconductor (S), paired electrons enter the N region, changing its
excitation spectrum. How drastic are these changes?
Recent studies~\cite{melsen,nazarov,golubov,spivak,belzig}
demonstrate that the answer
depends crucially on the type of dynamics in the N region:
In the case of {\em integrable}\/ classical dynamics, the density of
states
(DoS) of excitations is suppressed at low energies
and vanishes nearly linearly at the Fermi level.
Contrary, in N systems with {\em chaotic}\/ dynamics,
coupling to a superconductor produces a gap in the DoS
of the order of $\hbar/\tau_c$,
where $\tau_c$ is the typical time needed to establish contact
with the superconductor.

To illustrate this statement, consider a generic example
of a chaotic NS system: an SNS junction made of a disordered
conductor of size $L$ connected to the S terminals.
When the N metal is diffusive,
with the mean free path $l \ll L$, and sufficiently long,
with the Thouless energy $\ETh = \hbar D/L^2 \ll \Delta$
(here $D = v_Fl/3$ is the diffusion constant,
and $\Delta$ is the superconductive gap in the terminals),
then $\tau_c$ is given
by the diffusion time across the N region, $\tau_c \sim \hbar/\ETh$.
Thus, the energy gap $\gap \sim \ETh$
develops in the DoS~\cite{golubov,spivak}.

However, all the results mentioned above are based either on
the quasiclassical theory of superconductivity and proximity
effect~\cite{Eilenberger,classics,Usadel} or on the mead-field
treatment of the random-matrix theory (RMT).
Although usually this is a good approximation,
it is interesting to check whether some effects,
which are beyond quasiclassics, may lead to qualitative
changes in the above picture.

In this Letter we show that, indeed,
mesoscopic fluctuations smear the hard gap in the quasiparticle
spectrum of dirty SNS junctions, producing a tail of the subgap
states with energies $E < \gap$. These low-lying states are due to
the existence of quasi-localized states~\cite{MK,FE} in the N part
of the junction which are weakly coupled to the S terminals,
thereby having a larger $\tau_c$.
The magnitude of this mesoscopic effect is controlled by the
dimensionless normal-state conductance $G \equiv h/e^2R_n$ of the
junction, and is small (except for the vicinity of $\gap$)
at $G \gg 1$.
A related problem was considered
recently in Ref.~\cite{vavilov}, where
a hypothesis of universality and some nontrivial results from
the RMT~\cite{matrix} were used to find
an exponentially small DoS below the mean-field gap
in the N dot weakly connected to a superconductor.
We will show that their conjecture holds
for sufficiently narrow junctions close to $\gap$.
Technically, our approach is completely different
from Ref.~\cite{vavilov}
as we employ the fully microscopic
method based on the supermatrix $\sigma$-model method~\cite{efetov}
for mixed NS systems~\cite{simons},
which is free from any assumption about the random-matrix universality.
We find a nontrivial instanton solution
to the saddle-point equations of the supermatrix $\sigma$-model and show
that it is responsible for the non-zero DoS at energies $E < \gap$.
Our method is similar to the one used in~\cite{EM} to calculate
the DoS in the quantum Hall regime.
The relevance of the nontrivial saddle point solutions for the formation
of the DoS tail in NS systems was pointed out in Ref.~\cite{simons}.

We consider an SNS junction with the N region being a rectangular
bar of size $L \times L_y \times L_z$, coupled to the S terminals
by the ideal contacts situated at $x =\pm L/2$.
We neglect superconductivity suppression in the bulk terminals,
provided that they are sufficiently large,
and assume zero superconductive phase difference between them.
The dimensionless conductance of the N region,
$G(L,L_y,L_z)=4\pi\nu DL_yL_z/L$,
where $\nu$ is the normal-metal DoS per a single projection of spin.

Our results can be summarized as follows.
The behavior of the subgap DoS $\dos$ depends on the relation between
$L_y, L_z$, and the effective transverse length $L_\perp(E)$,
with the latter scaling as
$L_\perp(E\to\gap) \sim L (1-E/\gap)^{-1/4}$ and
$L_\perp(E\ll\gap) \sim L$.
In the vicinity of the quasiclassical gap, at $E\to\gap$,
 we find an intermediate asymptotic behavior
$\ln\dos \sim - G(L,{\cal L}_y(E),{\cal L}_z(E)) (1-E/\gap)^{3/2}
\sim -(1-E/\gap)^{(6-d)/4}$,
where ${\cal L}_{y,z}(E) = \min(L_{y,z},L_\perp(E))$,
and $d$ is the effective dimensionality of the junction:
$d=0$ for narrow junctions with $L_y,L_z\ll L_\perp(E)$,
$d=1$ for wider junctions with $L_y \gg L_\perp(E) \gg L_z$,
and $d=2$ for films with $L_y,L_z\gg L_\perp(E)$.
In the low-energy limit, $E \ll \gap$,
the behavior of the DoS is log-normal in 0D:
$\ln\dos \sim -G \ln^{2}(\gap/E)$,
and is a power law in 1D: $\ln\dos \propto -\ln(\gap/E)$.
These results are similar to those for the
distribution of relaxation times in 1D and 2D diffusive
conductors~\cite{MK},
although with an important difference: the basic energy scale
in our case is given by $\gap$, whereas in Ref.~\cite{MK}
it was the level spacing $\delta =1/\nu V$.

As a warm-up,
we recall the standard quasiclassical approach~\cite{golubov,spivak}
to diffusive SNS junctions based on the Usadel equation~\cite{Usadel}
for the quasiclassical retarded Green function $\hat{G}^R(E,{\bf r})$.
For the latter we will assume the angular parametrization,
$\hat{G}^R = \tau_z\cos\theta + \tau_x\sin\theta\cos\varphi +
\tau_y\sin\theta\sin\varphi$, where $\tau_i$ are the Pauli
matrices in the Nambu space.
In the absence of the phase difference between the S terminals,
one can set $\varphi \equiv 0$, and the Usadel equation acquires the
form
(hereafter $\hbar=1$)
\be
  D\nabla^2\theta + 2i E \sin\theta=0,
  \qquad
  \theta\left(x =\pm\frac{L}{2}\right)=\frac{\pi}2.
\label{Usadel}
\ee
The boundary conditions at the free edges of the N film are
of von Neumann type.
The local density of states is given by
$\left<\rho_{\rm local}(\br,E)\right> = 2\nu\Re\cos\theta(\br)$.
The substitution $\theta(x) = \pi/2 + i\psi(x)$ leads to the real
equation for the function $\psi(x)$ with $\psi(\pm L/2)=0$.
This equation can be easily integrated yielding the relation
between $E$ and the magnitude of $\psi(x)$ at $x=0$:
\be
  \sqrt{\frac{E}{\ETh}} =
  \int\limits_0^{\psi(0)}\frac{d\psi}{\sqrt{\sinh\psi(0)-\sinh\psi}}.
\label{2}
\ee
Eq.~(\ref{2}) has real solutions for $\psi(0)$ only
for $E\leq\gap = c\ETh$, with $c=3.12$. At $E>\gap$, $\psi(0)$ becomes
complex
resulting in the square-root singularity~\cite{golubov} in the DoS
(here and below we provide results for the DoS integrated over
the whole N region, $\dos=\int d\br\left<\rho_{\rm
local}(\br,E)\right>$\ ):
\be
  \dos_{\rm quasicl.} = 2.30\: \delta^{-1} \sqrt{E/\gap-1}.
\label{rho1}
\ee
Below the mean field gap, at $E < \gap$, Eq.~(\ref{2}) has two real
solutions, $\psi_1(0)$ and $\psi_2(0)>\psi_1(0)$,
merging at $E=\gap$. They determine the corresponding solutions
$\theta_{1,2}(x) = \pi/2 + i\psi_{1,2}(x)$ to Eq.~(\ref{Usadel}).
Having real $\psi(x)$, both of them do not contribute
to the DoS at $E < \gap$.
Usually the solution $\psi_1(x)$ is chosen
by the continuity argument, as it obeys the natural condition
$\lim_{E \to 0}\psi_1(x,E) = 0$, while $\psi_2(x,E)$ diverges
in the limit of vanishing $E$. We will see however that it is this
second solution which is responsible for the finite DoS below
the quasiclassical gap.

In order to extend the quasiclassical solution and take into account
mesoscopic fluctuations, we will use the nonlinear supermatrix
$\sigma$-model similar to that derived in \cite{simons}.
As we are interested in the average density of states only, it is
sufficient to calculate the retarded single particle Green function.
The derivation of $\sigma$-model starts with representing $\hat{G}^R$
in terms of the functional integral:
\begin{eqnarray}
  &\ds
  \hat{G}^R(\br,\br',E) =
    -\frac{i}{2}\int\phi_{\rm F}(\br)\phi^+_{\rm F}(\br')
    e^{-S[\phi]}D\phi^*D\phi ,
\\ \nonumber
  &\ds
  S[\phi] = \int d\br\; \phi^+(\br)(E+i0 - \hat{\cal H})\phi(\br).
\end{eqnarray}
In this expression $\phi$ is the 4-component superfield consisting of
commuting (complex) and anticommuting (Grassmann) parts.
The Hamiltonian $\hat{\cal H}$ is a matrix in the Nambu-Gor'kov (N)
space:
\be
  \hat{\cal H}
  = \tz\left[\frac {{\bf p}^2}{2m}-E_F+U(\br) \right]+\Delta(\br)\tx ,
\ee
where $\Delta(\br)=\Delta \theta(|x|-L/2)$, and $U(\br)$ is the
random potential.
In the absence of magnetic field, one has to double the field space
in order to account the time-reversal symmetry,
introducing the ``time-reversal'' (TR)
space according to $\Phi = ( \phi, i\ty\phi^* )^T/\sqrt{2}$.
This definition coincides with the one used in~\cite{skf};
it differs from the notations of~\cite{efetov,simons}
by the factor $i\ty$ in the $\phi^*$ sector.
Pauli matrices operating in the TR space are denoted by $\sigma_i$.

After this definitions the derivation of $\sigma$-model is
straightforward \cite{efetov,simons,skf}.
One has to carry out
(i)~averaging over the random potential with the correlator
$\left< U(\br)U(\br')\right> = \delta(\br-\br')/2\pi\nu\tau$,
(ii)~Hubbard-Stratonovich transformation introducing an $8\times8$
matrix $Q$ acting in the product of FB, N, and TR spaces,
(iii)~expansion to the leading terms in $\nabla Q$, $E$, and $\Delta$.
The result is
\begin{eqnarray}
  &\ds
  \left<\rho_{\rm local}(\br,E)\right>
    = \frac {\nu}4 \Re\int DQ \str\{k\Lambda Q(\br)\}\, e^{-S[Q]},
\label{rho2}
\\
  &\ds
  S[Q] = \frac {\pi\nu}8 \!\int\! d\br\;
    \str \left[D(\nabla Q)^2 + 4iQ(i\tx\Delta + \Lambda E)\right].
\label{act}
\end{eqnarray}
Here $\Lambda=\sz\tz$, the matrix $Q=U^{-1}\Lambda U$ with the proper
set~\cite{simons} of matrices $U$ is subject to the condition
$Q=CQ^TC^T$,
where $C = -\tx\sz [(1+k)\sigma_x+(1-k)i\sigma_y]/2$,
and $k = \mathop{\rm diag}(1,-1)_{\rm FB}$.
The manifold of $Q$-matrices
can be parametrized by 8 real and 8 Grassmann variables.

The next step is to find the saddle point solutions
to the action (\ref{act}). We start from the simplest case
when the $Q$ matrix does not contain Grassmann variables.
Then it splits into the Fermi-Fermi (FF) and Bose-Bose (BB) sectors
which can be parametrized independently by four variables in each
sector:
\begin{mathletters}
\label{parametr}
\begin{eqnarray}
  Q_{\rm FF} &=& \tz\cos\theta_{\rm F}[\sz\cos k_{\rm F}
  + \sin k_{\rm F} (\sx\cos\chi_{\rm F} + \sy\sin\chi_{\rm F})]
\nonumber \\
  &\mathrel{+}& \sin\theta_{\rm F}(\tx\cos\vphi_{\rm F} +
\ty\sin\vphi_{\rm F}),
\\
  Q_{\rm BB} &=&
  [\sz\cos k_{\rm B}
    + \tz\sin k_{\rm B} (\sx \cos\chi_{\rm B} + \sy \sin\chi_{\rm B})]
\nonumber \\
  &\mathrel{\times}&
  [\tz\cos\theta_{\rm B}
    + \sz\sin\theta_{\rm B} (\tx\cos\vphi_{\rm B} + \ty\sin\vphi_{\rm
B})].
\end{eqnarray}
\end{mathletters}%
In terms of the angular parametrization (\ref{parametr}),
the action (\ref{act}) acquires the form
$S = (\pi\nu/2) \int d\br ({\cal L}_{\rm FF}-{\cal L}_{\rm BB})$,
where
\begin{eqnarray}
{\cal L}_{\rm FF} &=& D\left[ (\nabla\theta_{\rm F})^2 \right.
+       \sin^2\theta_{\rm F}(\nabla\vphi_{\rm F})^2
\nonumber \\
&\mathrel{+}& \cos^2\theta_{\rm F}(\nabla k_{\rm F})^2
+       \left.\cos^2\theta_{\rm F}\sin^2k_{\rm F}(\nabla\chi_{\rm
F})^2\right]
\nonumber \\
&\mathrel{+}& 4iE \cos\theta_{\rm F} \cos k_{\rm F}
-       4\Delta \sin\theta_{\rm F} \cos\vphi_{\rm F},
\nonumber \\
{\cal L}_{\rm BB} &=& D\left[(\nabla\theta_{\rm B})^2\right.
+       \sin^2\theta_{\rm B}(\nabla\vphi_{\rm B})^2
\nonumber \\
&\mathrel{+}& (\nabla k_{\rm B})^2
+       \left. \sin^2k_{\rm B}(\nabla\chi_{\rm B})^2 \right]
\nonumber \\
&\mathrel{+}& 4iE \cos\theta_{\rm B} \cos k_{\rm B}
-       4\Delta \sin\theta_{\rm B} \cos k_{\rm B} \cos\vphi_{\rm B} .
\nonumber
\end{eqnarray}
The variables $\theta_{\rm F,B}$ and $\vphi_{\rm F,B}$ coincide
(at $k_{\rm F,B}=0$) with the standard Usadel angles, whereas
their counterparts $k_{\rm F,B}$ and $\chi_{\rm F,B}$ are new
ingredients of the field theory.

Minimization of the action for a uniform superconductor
at $E \ll \Delta$ gives $\theta_{\rm F,B}=\pi/2$ and $k_{\rm F,B}=0$,
that provides the boundary conditions for $Q$ in the N region.
In the absence of the phase difference between the S terminals,
one obtains $\varphi_{\rm F,B}=k_{\rm F}=0$ and $\chi_{\rm F,B}=\const$
at the saddle point in the N part of the structure.
Introducing new variables $\alpha_{\rm B}=\theta_{\rm B}+k_{\rm B}$
and $\beta_{\rm B}=\theta_{\rm B}-k_{\rm B}$ in the BB sector,
one obtains for the saddle-point action:
\begin{eqnarray}
  & \ds
  S[\theta_{\rm F},\alpha_{\rm B},\beta_{\rm B}] =
    2S_0[\theta_{\rm F}] - S_0[\alpha_{\rm B}] - S_0[\beta_{\rm B}] ,
\label{S}
\\
  & \ds
  S_0[\theta] =
    \frac{\pi\nu}4 \int d\br \,
    \left[D(\nabla\theta)^2 + 4iE\cos\theta \right] .
\label{S0}
\end{eqnarray}
Varying with respect to $\theta$, one recovers Eq.~(\ref{Usadel})
as the saddle-point equation for the action~(\ref{S0}).

The Usadel equation (\ref{Usadel}) possesses,
apart from the $x$-dependent solutions discussed above,
solutions which depend on the transverse ($y,z$) coordinates.
The role of the transverse dimensions will be discussed later,
while now we will consider the 0D case, relevant for sufficiently
narrow junctions with $L_x,L_y\ll L_\perp(E)$.
Then, according to the previous analysis, the Usadel equation
has two solutions, $\theta_1(x)$ and $\theta_2(x)$.
Therefore, the full action (\ref{S}) has in total 8 different
saddle point solutions:
$(\theta_{\rm F},\alpha_{\rm B},\beta_{\rm B}) =
(\theta_i,\theta_j,\theta_k)$, with $i$, $j$, $k=1,2$,
that will be refered to as $(i,j,k)$.
However, only 4 of them with $\theta_{\rm F}=\theta_1$ can be
reached by a proper deformation of the integration contour.

The simplest is the {\em supersymmetric} saddle point (1,1,1).
In this case, Gaussian integrations over commuting and anticommuting
variables near it
cancel each other, and the contribution to $\dos$
reduces to the form (\ref{rho1}) with the vanishing DoS below $\gap$.
Thus, the saddle-point approximation for the $\sigma$-model
(\ref{act}) restricted to the supersymmetric saddle point
is equivalent to the quasiclassical treatment based on the Usadel
equation (\ref{Usadel}).

To get a non-zero DoS below $\gap$ it is necessary to take
saddle points with {\em broken supersymmetry} into account.
Such a solution with the lowest action is given by (1,1,2)
[actually, a whole degenerate
family of the saddle points, and, in particular, (1,2,1),
can be obtained from it by rotation on the angle $\chi_B \in [0,2\pi)$].
The key point is that Gaussian fluctuations near this saddle point
have a negative eigenvalue which leads to an additional
imaginary unity in the preexponent and, consequently,
to the nonzero contribution to the DoS.
This contribution is suppressed by the factor $e^{-\Delta S}$,
where $\Delta S=S_0[\theta_1]-S_0[\theta_2]>0$ is the difference
between the actions of the solutions $\theta_{1,2}(x)$.
Finally, the saddle point (1,2,2) has the action
$2\Delta S$ and its contribution can be disregarded at $\Delta S\gg1$.
Thus, the subgap DoS can be estimated with exponential
accuracy as $\dos \sim \delta^{-1} e^{-\Delta S(E)}$.
Below we will calculate $\dos$ in the limiting cases
$E\to\gap$ and $E\ll\gap$.

The solutions $\theta_1(x)$ and $\theta_2(x)$ merge at $\gap$.
This fact can be used to find the asymptotically exact result for
$\dos$ in the energy range $G^{-2/3} \ll 1-E/\gap \ll 1$.
Let us start with the supersymmetric saddle point (1,1,1)
and look at fluctuations around it.
Almost all of them are hard, having a mass of the order of (or larger
than) $\gap$. There are only 8 (corresponding to 4 commuting and 4
anticommuting variables) soft modes whose mass vanishes at $E=\gap$.
Half of them transform the saddle point (1,1,1) to the instanton
(1,1,2),
and the other half transform it to the instanton (1,2,2).
Below we will consider the case when $\Delta S\gg1$
that allows to disregard the contribution of the instanton (1,2,2)
and to take Gaussian integrals over the corresponding soft and all
hard fluctuations.
As a result, we end up with 2 commuting ($q$ and $\chi_{\rm B}$)
and 2 Grassmann ($\zeta$ and $\xi$) variables parametrizing
the relevant soft degrees of freedom in the matrix
$Q=e^{-W^c/2}e^{-W^a/2}\Lambda e^{W^a/2}e^{W^c/2}$.
Here the matrix $W^a$ contains anticommuting variables:
$W^a_\fb = (if_0(x)/4)
[(\zeta+\xi)(i\ty+\tz\sx) + (\zeta-\xi)(i\ty\sz+\tz i\sy)]$,
$W^a_{\rm BF} = \tx\sx (W^a_{\rm FB})^T \tx i\sy$,
while in the absence of $W^a$, $Q$ reduces to the form (\ref{parametr})
with $\theta_{\rm F}=\beta_{\rm B}=\theta_1(x)$,
$\alpha_{\rm B}=\theta_1(x)+iqf_0(x)$,
$k_{\rm F}=\vphi_{\rm F}=\vphi_{\rm B}=0$.
The function $f_0(x)$ is the normalized difference
$\delta\psi(x) = \psi_2(x) - \psi_1(x)$ at $\gap$:
$f_0(x) = \lim_{E \to \gap} \delta\psi(x)/||\delta\psi(x)||$,
where $||F(x)||^2=(1/L)\int_{-L/2}^{L/2} F^2(x) dx$.
Evaluating the action, integrating over the cyclic angle $\chi_B$,
and performing the $x$-integration, one obtains
\begin{eqnarray}
  & \ds
  S = \K \left[ \sqrt{\eps} q^2-\frac{q^3}3+\zeta\xi(2\sqrt{\eps}-q)
\right],
\label{s0d}
\\
  & \ds
  \K = \frac{\pi c_2\gap}{2\delta} = \frac{c c_2}8 G, \qquad
  \eps = \frac {2c_1}{c_2} \frac{\gap-E}{\gap},
\label{eps}
\end{eqnarray}
where $c_n=\int_{-L/2}^{L/2} \cosh\psi_0(x)f_0^{2n-1}(x)\,dx/L$,
and $\psi_0(x)=\psi_{1,2}(x,\gap)$;
$c_1 = 1.15$, $c_2 = 0.88$.
The measure of integration (including the Berezenian)
is given by $DQ=q\,dq d\zeta d\xi$.
We take the contour of integration over $q$ along the imaginary axis to
get
the convergent integral. Finally, substituting
$\int\str(k\Lambda Q)dV = 2i c_1 V (4\sqrt{\eps}-q)$
into Eq.~(\ref{rho2}), we arrive at the one-instanton correction
to the quasiclassical result (\ref{rho1}):
\be
  \dos
  = \frac{c_1}{2\delta} \Im \int\limits_0^{i\infty+0}
    \exp\left[ -\K \left( \sqrt{\eps} q^2-\frac{q^3}3 \right) \right] dq
.
\label{int_dq}
\ee
Eq.~(\ref{int_dq}) is valid provided that $\eps\gg\K^{-2/3}$
when the contribution of the instanton (1,2,2) can be neglected.
Then the integral over $q$ can be calculated by the saddle point method
yielding
at $\K^{-2/3}\ll\eps\ll 1$ in the 0D case:
\be
  \left<\rho(E\to\gap)\right>_{\rm 0D} =
    \frac{c_1}{4\delta} \sqrt{\frac{\pi}{\K\sqrt{\eps}}}
    \exp\left( -\frac43 \K\eps^{3/2} \right) .
\label{0D1}
\ee
This result can be generalized for a normal dot of an arbitrary
shape coupled to a superconductor (cf.~\cite{melsen,vavilov}),
provided that the numbers $c$, $c_n$ are defined with the use of the
exact solutions $\theta_{1,2}(\br)$ of the Usadel equation in a
given geometry: $c_n=(1/V)\int\cosh\psi_0(\br)f_0^{2n-1}(\br)\,d\br$.
The functional form of the result (\ref{0D1}) coincides with
the RMT conjecture of Ref.~\cite{vavilov}.
An exact correspondence with the RMT prediction~\cite{vavilov}
is expected in the case of weakly transparent NS interfaces.

Coming back to the planar SNS junction, we turn to
the limit of small energies, $E \ll \gap$. Here
one has $\psi_1(x) \approx 0$ and $\psi_2(x) \approx A(1-2|x|/L)$,
where, according to Eq.~(\ref{2}), $A \approx \ln(\gap/E)$
(cf.\ Ref.~\cite{MK}).
Thus, the action of the instanton (1,1,2) becomes
$\Delta S(E) \approx - S_0[\theta_2] = \pi\ETh A^2/\delta$,
and the result for DoS reads
\be
  \left<\rho(E\ll\gap)\right>_{\rm 0D} \sim \frac 1{\delta}
    \exp\left( -\frac{G}4 \ln^2\frac{\gap}{E} \right).
\label{0D2}
\ee

Now let us consider the role of the saddle-point solutions
which depend on the transverse coordinates $y,z$.
At $E\to\gap$, one has to retain only soft modes
associated with the instanton (1,1,2). As a result,
the action (\ref{s0d}) acquires a gradient term:
\be
  S = \K \int \frac{dy}{L_y} \frac{dz}{L_z}
    \left(
      \frac{L^2}{4cc_2}(\nabla_\perp q)^2 + \sqrt{\eps}q^2-\frac{q^3}3
    \right),
\label{Syz}
\ee
where the Grassmann variables are discarded as we are not
interested in the preexponent.
Comparing the first and the second term in Eq.~(\ref{Syz}),
one extracts the characteristic transverse scale
$L_\perp(E) = L(cc_2)^{-1/2}\eps^{-1/4} \sim L\eps^{-1/4}$,
which determines the effective dimensionality of the system.
If $L_y$ or $L_z$ is shorter than $L_\perp(E)$,
then it costs too much energy to have gradients in that direction,
and the corresponding dimension ``freezes out''.
The 0D case considered above referred to the limit
$L_y,L_z\ll L_\perp(E)$.
Otherwise, an instanton will appear in the transverse direction
to minimize the total action.
In the 1D case ($L_y \gg L_\perp(E) \gg L_z$),
the action~(\ref{Syz}) achieves its stationary point at
$q(y) = 3\sqrt{\eps}\cosh^{-2}(y/L_\perp)$,
leading to
\be
  \left<\rho(E\to\gap)\right>_{\rm 1D} \sim \frac 1{\delta}
  \exp\left(-\frac{12\pi}{5}\sqrt{cc_2} \, \nu DL_z \,
\eps^{5/4}\right).
\label{1D1}
\ee
Analogously, in the 2D case ($L_y, L_z \gg L_\perp(E)$),
the quasiparticle DoS tail has the form
\be
  \left<\rho(E\to\gap)\right>_{\rm 2D} \sim \frac 1{\delta}
  \exp\left( -24.4\: \nu DL \, \eps \right).
\label{2D1}
\ee
The length $L_\perp(E)$ diverges at $\gap$, indicating that
any junction becomes effectively 0D
close to
the quasiclassical gap.
However, different parts of the DoS tail may
exhibit different exponents, from 1 to 3/2
[cf.\ Eqs.~(\ref{0D1}), (\ref{1D1}), and (\ref{2D1})],
with some cross-over in between.

The value of $L_\perp(E)$ is getting shorter as $E$ decreases,
and becomes of the order of $L$ at $E\leq\gap$.
The solution of the 1D problem at $E \ll \gap$
can be found following Ref.~\cite{MK}.
The function $\psi_2(x,y)$ has a sharp peak at the
center of the instanton
and with the logarithmic accuracy is given by
$\psi_2(x,y)=-4\ln(2\sqrt{x^2+y^2}/L)$.
The result for the DoS then reads
\be
  \left<\rho(E\ll\gap)\right>_{\rm 1D} \sim
  \frac 1{\delta}
    \left(\frac{E}{\gap}\right)^{4\pi^2\, \nu D L_z} .
\label{1D2}
\ee
The DoS at $E\ll\gap$ in the 2D case can be calculated
in the same way, and appears to be vanishingly small.

To conclude, we have shown that mesoscopic fluctuations
smear the quasiclassical gap in the DoS of a diffusive SNS junction.
The tail in the DoS is due to the states anomalously localized in the
N part of the junction and weakly coupled to the S terminals.
Technically, the tail is described by instantons with broken
supersymmetry.

At the final stage of the preparation of this manuscript
we became aware of Ref.~\cite{Simons}
where a similar instanton solution is found for the problem
of tail states below $E_g$ in a superconductor with magnetic impurities.

We are grateful to C. W. J. Beenakker, D. A. Ivanov, A. D. Mirlin,
Yu. V. Nazarov and A. V. Shytov for useful discussions.
This research was supported by the NWO-Russia collaboration grant,
Swiss NSF-Russia collaboration grant, RFBR grant 98-02-16252,
and by the Russian Ministry of Science within the project
``Mesoscopic electron systems for quantum computing".


\end{multicols}

\begin{references}


\bibitem{melsen} J. A. Melsen et al, Europhys. Lett. {\bf 35}, 7 (1996);
Physica Scripta {\bf 69}, 223 (1997).

\bibitem{nazarov} A. Lodder and Yu. V. Nazarov,
Phys. Rev. B {\bf 58}, 5783 (1998).

\bibitem{golubov} A. A. Golubov and M. Yu. Kupriyanov, Sov. Phys. JETP
{\bf 69}, 805 (1989).

\bibitem{spivak} F. Zhou et al, J. Low Temp. Phys. {\bf 110}, 841
(1998).

\bibitem{belzig} S. Pilgram, W. Belzig and C. Bruder, cond-mat/0006222.

\bibitem{Eilenberger} G. Eilenberger, Z. Phys. {\bf 214}, 195 (1968).

\bibitem{classics} A. I. Larkin and Yu. N. Ovchinnikov,
Sov. Phys. JETP {\bf 26}, 1200 (1968).

\bibitem{Usadel} K. Usadel, Phys. Rev. Lett. {\bf 25}, 507 (1970).

\bibitem{MK} B. A. Muzykantskii and D. E. Khmelnitskii,
Phys. Rev. B {\bf 51}, 5480 (1995).

\bibitem{FE} V. I. Fal'ko and K. B. Efetov,
Europhys. Lett. {\bf 32}, 627 (1995).

\bibitem{vavilov} M. Vavilov et al, cond-mat/0006375.

\bibitem{matrix} C. A. Tracy and H. Widom, Comm. Math. Phys. {\bf 159},
151 (1994); {\bf 177}, 727 (1996).

\bibitem{efetov} K. B. Efetov, {\it Supersymmetry in Disorder and Chaos}
(Cambrigde University Press, New York, 1997).

\bibitem{simons} A. Altland, B. D. Simons, D. Taras-Semchuk,
Adv. Phys. {\bf 49}, 321 (2000).

\bibitem{EM} K. B. Efetov and V. G. Marikhin,
Phys. Rev. B {\bf 40}, 12126 (1989).

\bibitem{skf} M. A. Skvortsov, V. E. Kravtsov, M. V. Feigel'man,
JETP Lett. {\bf 68}, 84 (1998).

\bibitem{Simons} A. Lamacraft, B. D. Simons,
Phys. Rev. Lett. {\bf 85}, 4783 (2000).


\end{references}
\end{document}